# Bow-tie topological features of metabolic networks and the functional significance

ZHAO JING [1,2,4], TAO LIN [2], YU HONG[2], LUO JIAN-HUA[1,2], CAO Z. W. [2] and LI YIXUE[2,3,1]

[1] School of Life Sciences & Technology, Shanghai Jiao Tong University, Shanghai 200240, China.
[2] Shanghai Center for Bioinformation and Technology, Shanghai 200235, China
[3] Shanghai Institutes for Biological Sciences, Chinese Academy of Sciences, Shanghai 200031, China
[4] Department of mathematics, Logistical Engineering University, Chongqing 400016, China
Correspondence should be addressed to Cao Z W and Li Yi-Xue (email:zwcao@scbit.org, yxli@sibs.ac.cn )

**Abstract**  Exploring the structural topology of genome-based large-scale metabolic network is essential for investigating possible relations between structure and functionality. Visualization would be helpful for obtaining immediate information about structural organization. In this work, metabolic networks of 75 organisms were investigated from a topological point of view. A spread bow-tie model was proposed to give a clear visualization of the bow-tie structure for metabolic networks. The revealed topological pattern helps to design more efficient algorithm specifically for metabolic networks. This coarse-grained graph also visualizes the vulnerable connections in the network, and thus could have important implication for disease studies and drug target identifications. In addition, analysis on the reciprocal links and main cores in the GSC part of bow-tie also reveals that the bow-tie structure of metabolic networks has its own intrinsic and significant features which are significantly different from those of random networks.

**Keywords:**   bioinformatics; metabolic network; bow-tie; random network.

As the results of various genome projects, the genome sequences of many organisms are available and the organism specific metabolic networks can be faithfully reconstructed from genome information [1-6]. Thus the prediction of function from the metabolic networks has become an essential step in the post-genomic era[4, 7-11]. However, before it is possible to investigate the potential relationship between structure and functionality of a metabolic network, it is necessary to study how the metabolic networks are actually constructed.

In recent years there has been a strongly increasing passion in investigating the intricate structures of link topologies in metabolic networks [12-21]. Many researchers have applied methods from graph theory to treat the metabolic networks as a directed graph consisting of nodes denoting metabolites connected by directed arcs representing reactions. An important finding is that metabolic networks, as well as other real-world complex networks, have topologies that differ markedly from those found in simple randomly connected networks [22], which suggests that their non-random structures could imply significant organizing principles of metabolic networks.

From the computational analysis of genome-based metabolic networks of 65 organisms, Ma and Zeng discovered that the global metabolic network is organized in the form of a bow-tie[15]. On the other hand, from the view of material flows and information flows in the metabolic system, Csete and Doyle described metabolism as several nested bow-ties[23]. It was pointed out that bow-tie architectures facilitate robust biologic function, and based on their design, also have inherent but predictable fragilities. They concluded that large-scale organizational frameworks such as the bow-tie are necessary starting points for higher-resolution modeling of complex biologic processes[23].

As the bow-tie structure of networks has drawn so much attention [15, 23-26], clear visualization method of this architecture is being expected because of the large number of nodes and arcs, while GSC, the central part of bow-tie, deserves more detailed exploration. The decomposition of a network into *k*-cores [27] allows us to classify nodes simultaneously by both their connectivity and central placement in the network. Having been successfully applied to analyze protein interaction networks [28], this approach could help to probe the topological feature of the GSC part. In addition, metabolic networks include both irreversible reactions and reversible ones, which have different functions in the regulation of metabolic process. Thus the distribution feature of reversible reactions in metabolic networks may afford some useful insight into metabolic regulation, while the reciprocity metric has been defined to measure this feature of directed networks [29].

In this work, 75 metabolic networks were constructed from organisms including 8 eukaryote, 56 bacteria and 11 archaea. A spread bow-tie model, the modified bow-tie model based on a coarse-grained graph, of the metabolic network is proposed to visualize the global bow-tie structure. Then the reciprocal feature and the main core are explored in details for the complex GSC part of the bow-tie. To mine the inherent topology of metabolic networks, the features of *E.coli* network were then compared with those of the properly randomized counterparts that preserve the linkage degree of each node and the total number of directed and bi-directed arcs.

## 1. Materials and methods

*1.1 Data preparation and network reconstruction*

In this study, the metabolic data were extracted from the database developed by Ma and Zeng based on the Kyoto Encyclopedia of Genes and Genomes (KEGG)[5]. In this database, the information concerning the reversible reactions was specified. In addition, some small molecules, such as adenosine triphosphate (ATP), adenosine diphosphate (ADP), nicotinamide adenine dinucleotide (NAD) and $H_2O$, are normally used as carriers for transferring electrons or certain functional groups and participate in many reactions, but typically not participating in product formation. Therefore, in order to reflect biologically relevant transformations of substrates, these kinds of small molecules, as well as their connections were manually excluded from the database when no products were formed from them. It should be noted that this method of exclusion is not determined by compounds, but by the reaction. For example, glutamate (GLU) and 2-oxoglutarate (AKG) are currency metabolites for transferring amino groups in many reactions, but in the following reaction:

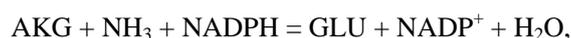
$$AKG + NH_3 + NADPH = GLU + NADP^+ + H_2O,$$

AKG participates in producing GLU, i.e. they are primary metabolites. Hence the connections through them should be considered. A metabolic network reconstructed from this database is represented by a directed graph whose nodes correspond to metabolites and whose arcs correspond to reactions between these metabolites, in which irreversible reactions are presented as directed arcs , and reversible ones as bi-directed arcs. For example, the irreversible reaction,

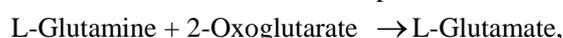
$$\text{L-Glutamine} + \text{2-Oxoglutarate} \rightarrow \text{L-Glutamate},$$

corresponds to two directed arcs, i.e. L-Glutamine $\rightarrow$ L-Glutamate and 2-Oxoglutarate

→L-Glutamate.

*1.2 Bow-tie structure*

A directed graph with bow-tie structure consists of four parts: giant strong component (GSC), substrate subset (S), product subset (P) and isolated subset (IS) [15]. The GSC is the biggest of all strongly connected components and is much larger than all the other ones, while a strongly connected component is defined as the largest cluster of nodes within which any pair of nodes is mutually reachable from each other [30]. S consists of nodes that can reach the GSC but cannot be reached from it, while P consists of nodes that are accessible from the GSC, but do not link back to it. The IS contains nodes that cannot reach the GSC, and cannot be reached from it.

*1.3 Spread bow-tie model*

The spread bow-tie model is proposed to clearly visualize the bow-tie structure of the metabolic network. It includes three steps as follows:

(1) Reduce the number of nodes and arcs on the condition of retaining the information flow of the network through a coarse-grained processing: each strongly connected component is treated as a cluster and is shrunk into a single node, while all the arcs from cluster *i* to cluster *j* are shrunk into a single arc.

(2) Further divide the nodes in the substrate (S), product (P) and isolated subset (IS) into more specific subsets depending on their connectivity pattern and biological features.

For every node *v* in set S and P of the coarse-grained graph, four parameters are used to depict its connectivity feature as follows:

$d^-(v)$: in-degree of node *v*, i.e., the number of links to node *v*;

$d^+(v)$: out-degree of node *v*, i.e., the number of links from node *v* to other nodes;

$d(v, G)$: the directed distance from node *v* to the GSC(G), *i.e.* the number of connections in the shortest directed path from node *v* to the GSC;

$d(G, v)$: the directed distance from the GSC(G) to node *v*, *i.e.* the number of connections in the shortest directed path from the GSC to node *v*.

According to these parameters, S and P are respectively divided into four subsets as follows:

S = S1 ∪ S2 ∪ S3 ∪ S4
P = P1 ∪ P2 ∪ P3 ∪ P4

where,

$S1 = \{v | d^-(v) = 0, d(v, G) > 1\}$

$S2 = \{v | d^-(v) > 0, d(v, G) > 1\}$

$S3 = \{v | d^-(v) > 0, d(v, G) = 1\}$

$S4 = \{v | d^-(v) = 0, d(v, G) = 1\}$

$P1 = \{v | d^+(v) = 0, d(G, v) = 1\}$

$$P2=\{v|d^+(v)>0, d(G,v)=1\}$$

$$P3=\{v|d^+(v)>0, d(G,v)>1\}$$

$$P4=\{v|d^+(v)=0, d(G,v)>1\}$$

Then the nodes in the isolated subset (IS) are divided into S-IS and P-IS according to whether they are connected with the S or P component.

(3) Draw the spread bow-tie by the graph analysis software Pajek [31].

First, draw the graph using the command "draw-partition-vector", in which "partition" and "vector" are vectors corresponding to the belongingness of subsets and the size of the nodes respectively. Then in the graph window, choose to locate the nodes in the same subset into one layer and arrange the layers in y-direction (command: "layers-In y direction"). Finally, adjust the location of a minority of nodes manually to get a clearer layout.

*1.4 Reciprocity metric*

In metabolic networks, irreversible reactions and reversible ones are represented by directed arcs and bi-directed arcs (or called reciprocal arcs), respectively. The reciprocity metric was defined to measure the link reciprocity of directed networks [29] as the correlation coefficient between the entries of the adjacency matrix of a directed graph:

$$\rho = \frac{\sum_{i \neq j}(a_{ij} - \bar{a})(a_{ji} - \bar{a})}{\sum_{i \neq j}(a_{ij} - \bar{a})^2}$$

where, the average value $\bar{a} = \frac{\sum_{i \neq j} a_{ij}}{N(N-1)} = \frac{L}{N(N-1)}$ (L and N are the number of total arcs and nodes in the network, respectively) measures the ratio of observed to possible directed arcs. In general, directed networks range between the two extremes of a purely bidirectional one (such as the Internet, where information always travels both ways along computer cables) and of a purely unidirectional one (such as citation networks, where recent papers can cite earlier ones while the opposite cannot occur). The range of $\rho$ is [-1,1], and $\rho$ equals to -1 and 1 for perfectly anti-reciprocal networks (purely unidirectional ones) and perfectly reciprocal networks (purely bidirectional ones) respectively. The reciprocity metric can be used to compare the degree of link reciprocity between graphs with different nodes and arcs.

*1.5 k-core of graphs*

In graph theory, a *k*-core H of graph G is a maximal subgraph of G such that every node in H has at least *k* links as Fig. 1. shows [27]. The core of maximum order, which is the most densely connected part of a graph, is also called main core.

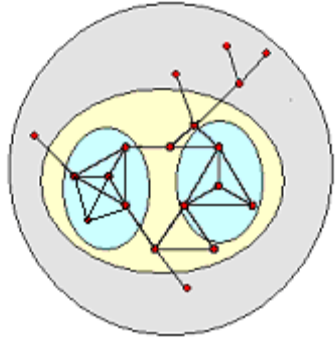

Fig. 1. 1, 2 and 3-core. Two basic properties of cores: first, cores may be disconnected subgraphs; second, cores are nested: for *i>j*, an *i*-core is a subgraph of a *j*-core of the same graph.

*1.6 Clustering coefficient*

The clustering coefficient of node *v* measures the extent that its neighbours are also linked together, i.e. to form a clique:

$$CC(v) = \frac{2|N(v)|}{d(v)(d(v)-1)},$$

where $|N(v)|$ denotes the number of links between neighbours of node *v*, *d(v)* is the degree of node *v*. The value of *CC (v)* is between 0 and 1. To some extent, the clustering coefficient of a network, i.e., the average of *CC (v)* over all *v*, could reflect the cliquishness of the network[32].

*1.7 Method to compare a real metabolic network with randomized ones*

Following the scheme of Maslov *et al.*[33], we apply Z-score to quantify the difference between a real metabolic network and its randomized counterparts:

$$Z = \frac{P - \overline{P_r}}{\Delta P_r},$$

where P is the graph metric in the real network, $\overline{P_r}$ and $\Delta P_r$ are the mean and standard deviation of the corresponding graph metric in the randomized ensemble.

## 2. Results and discussion

*2.1 Spread bow-tie of E.coli metabolic network*

In this section, we apply the spread bow-tie model to obtain a clear visualization of the bow-tie topology of *E.coli* metabolic network.

The metabolic network of *E. coli* K-12 MG1655 consists of 934 nodes and 1437 arcs. The largest connected part of this network embraces 575 nodes and its topology exhibits a bow-tie architecture, with 234, 85, 177 and 79 nodes in the GSC, S, P and IS part, respectively. According to the spread bow-tie model, this biggest connected cluster of 575 nodes is first shrunk into a coarse-grained graph of 215 nodes. Then the nodes of the coarse-grained graph are divided into 11 subsets as shown in Table 1. Finally, the modified bow-tie structure of the coarse-grained graph for the *E.coli* network looks like a butterfly with its wings spread as shown in Fig.2. That is why this model is called spread bow-tie model.

Table 1    The number of nodes in each subset of different bow-tie models

|   |   | Bow-tie of original graph | Bow-tie of the coarse-grained graph | Spread bow-tie of the coarse-grained graph |
|---|---|---|---|---|
| GSC | | 234 | 1 | 1 |
| S | S1 | 85 | 63 | 12 |
|   | S2 | | | 7 |
|   | S3 | | | 7 |
|   | S4 | | | 37 |
| P | P1 | 177 | 92 | 26 |
|   | P2 | | | 20 |
|   | P3 | | | 21 |
|   | P4 | | | 25 |
| IS | S-IS | 79 | 49 | 23 |
|   | P-IS | | | 26 |
| Total | | 575 | 215 | 215 |

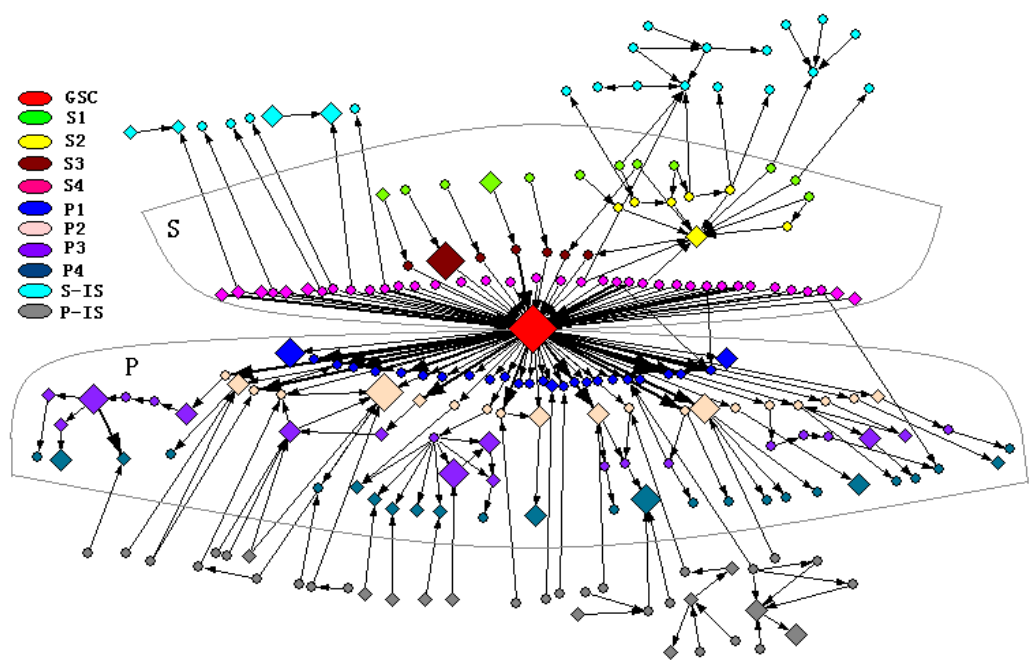

Fig. 2.   Spread bow-tie structure of the coarse-grained graph for *E.coli* metabolic network. The nodes in different colors and different layers belong to different subsets. The diamond represents strong component with at least two nodes, whereas the circle represents that with only one node. A bigger diamond corresponds to a strong component with more nodes. The width of an arc is proportional to the number of reactions between the two corresponding clusters. This graph is drawn by Pajek.

From the spread bow-tie structure, it is noticeable that the metabolic flow is highly branched, with the GSC as the metabolites converting hub. There are seldom cross-links between the branches, while most nodes in S, P and IS part are connected by some single linear paths. It is counted that among the 90 S-GSC and GSC-P arcs, 67 ones are single ones (thin lines), taking the percentage of 75%. Thus most connections of S-GSC and GSC-P are single arcs. Understanding such structure feature may help to design more efficient network algorithms specifically for metabolic networks. For example, when designing an algorithm to decompose the metabolic network into modules[34], we decomposed the GSC first and then expand the partition of GSC to other parts by a "majority rule"[24]. The sub-networks got from this algorithm exhibit a highly modularized bow-tie topological pattern similar to that of the global metabolic networks, while these small bow-ties are hierarchically nested into larger ones and collectively integrated into a large metabolic network. Thus our decomposition reveals the specific organization pattern of metabolic networks.

In addition, the global role of each node in the whole metabolic network is indicated in this model. The definition of the 11 subsets not only represents the connectivity feature of nodes in graph but also have biological significance. Metabolites corresponding to nodes in S1 and S4 are essential nutrients which *E. coli* may directly takes in from the environment or medium. Those in P1 and P4 are end products that need to be discharged. While those in S2, S3, P2 and P3 are intermediate metabolites. Metabolites corresponding to nodes in S3 and S4 are direct substrates going to GSC. Those in S1 and S2 can produce metabolites in GSC by at least two reactions. Similarly, metabolites corresponding to nodes in P1 and P2 are direct products from the GSC, while those in P3 and P4 can be produced from GSC metabolites by more than one reaction.

Furthermore, since each node in one strong component can reach any other members in the same component, an outside node linked to this strong component is thus able to reach (or be reached from) any members of this component. Therefore, the reduction according to strong component is a reasonable method to simplify complicated metabolic network while keeps its information flow, and the resulting coarse-grained graph may be conceived as macro-level framework of the metabolic network. In this way, the original interleaving and complicated metabolic network has been summarized into a clearly branched spread bow-tie model, which helps to illustrate the global biological metabolic flow.

*2.2 Reciprocity of the GSC*

To study the distribution of reversible reactions in metabolic networks, we computed the reciprocity metric in the whole network and the GSC part respectively for 75 organisms in the database of Ma and Zeng. The results are listed in part I of supplementary material. The normal quartile plots for the reciprocity in the whole network and the GSC part are shown in Fig 3. Both of the two plots are nearly linear, which indicates that the reciprocity metric in the whole network and the GSC part of the 75 organisms obey normal distributions. By minimum variance unbiased estimate, we got the average expectation and the standard deviation of the normal distribution for the reciprocity metric in the whole network are 0.6324 and 0.0344 respectively, while those for the reciprocity in GSC part are 0.8575 and 0.0482 respectively. That is to say, the average reciprocity in GSC is much higher than that in the whole network, suggesting that metabolic networks concentrate a big fraction of reversible reactions into the GSC part. The different average expectation of normal distribution for reciprocity in the whole network and the GSC also suggests that the concentrative distribution of reciprocity structure is an intrinsic topological feature of metabolic networks.

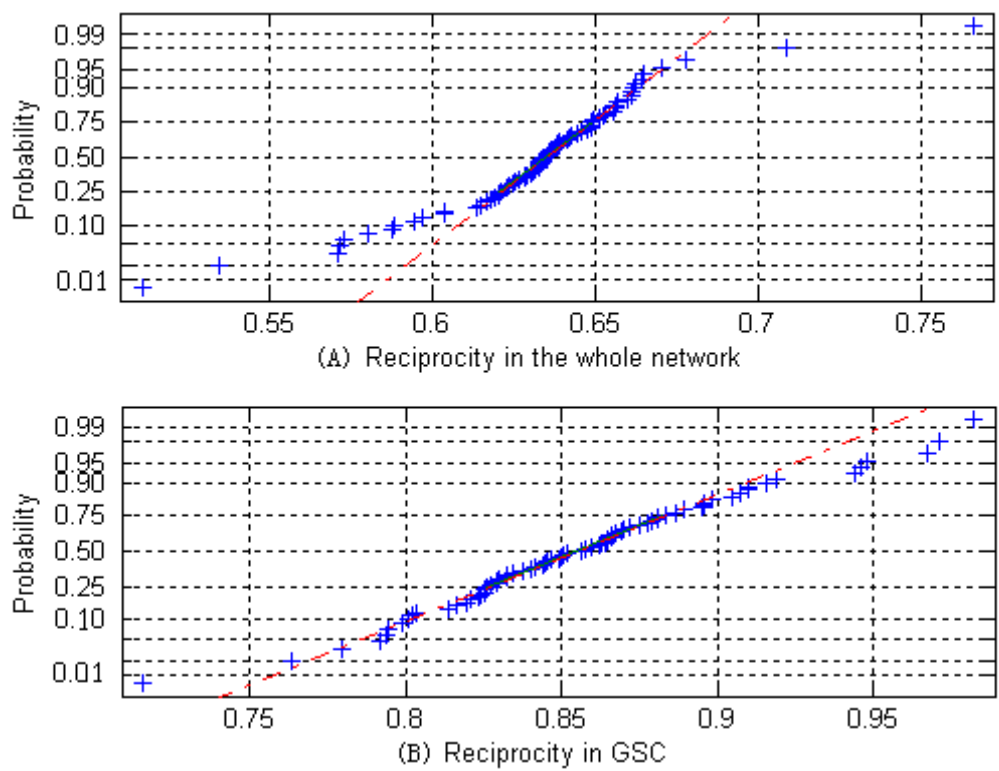

Fig. 3. Normal quartile plots for the reciprocity in the whole network and in the GSC of 75 organisms in the database of Ma and Zeng

*2.3 Main cores of the GSC*

Main core is the most densely connected part of a graph, thus could play important role in the network functionality. By detecting the connection density in the GSC, two 3-cores are found as main cores, which are bridged by two triangle nodes (2-Dehydro-3-deoxy-6-phospho-D-gluconate and Methylglyoxal) as Fig.4 shows.

When projecting to KEGG pathway[2, 3] (http://www.genome.jp/kegg/pathway.html), most reactions fall in carbohydrate metabolism, while only 6 metabolites participate in amino acid metabolism. Remarkably, the three major pathways –glycolysis, tricarboxylic acid (TCA) and pentose phosphate pathway take up most of the carbohydrate reactions in the main cores. Moreover, among the 12 precursors commonly used for biosynthesis[23], six ones are included in the main cores. This finding indicates that, as the essential and housekeeping processes in life, carbohydrate metabolism, especially main reactions in glycolysis, citrate cycle and pentose phosphate pathway are well protected and highly tolerant of random attacks.

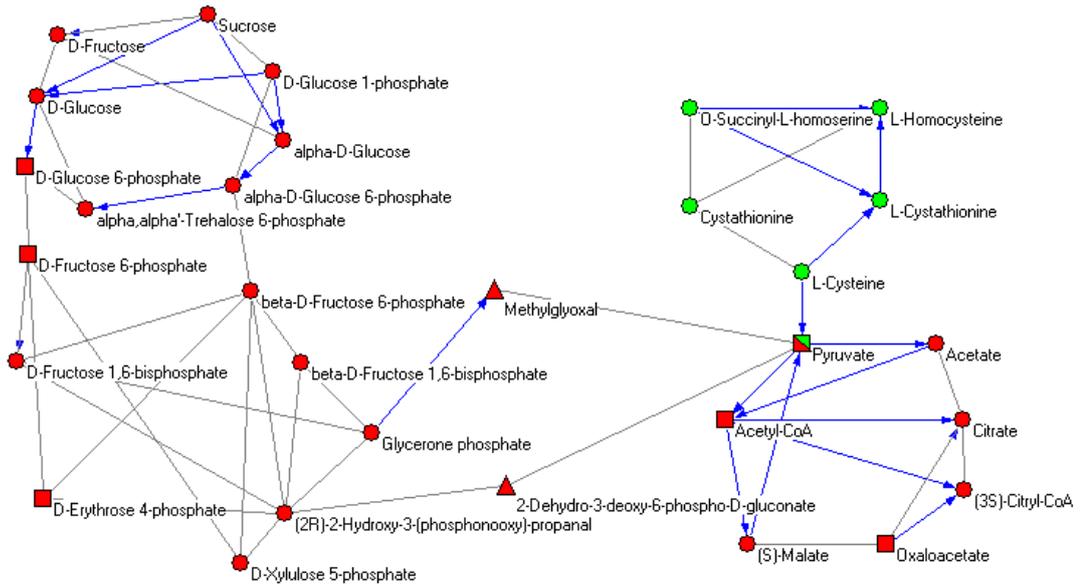

Fig.4.  Two main cores of the GSC, and the two metabolites (triangle nodes) directly connecting them. Red nodes participate in carbohydrate metabolism while green nodes include in amino acid metabolism. Six precursors (box nodes) are included in the main cores. For simplicity, reversible reactions are presented by undirected edges.

*2.4 Comparison between the E.coli network and an ensemble of randomly connected networks*

Maslov *et al.* proposed that the network under study should compared with its properly randomized version that preserves some of its low-level topological properties, such as degree sequence, while only statistically significant deviation of the topological property in question from that of its null model is meaningful [33, 35]. Hence, comparing metabolic network with randomized counterparts could reveal intrinsic difference between them.

In this study, we have known that the distribution of bi-directed arcs is an intrinsic topological feature of metabolic networks. Therefore, we construct the null model of metabolic networks as randomized networks that preserve not only the linkage degree of each node [33, 35] but also the total number of directed and bi-directed arcs of the metabolic network under study. The algorithm proposed in our another study is applied here to construct sixty random counterparts of the *E.coli* metabolic network [34]. The graph metrics of the 60 random networks are listed in Table S4 of the supplementary and the comparison with the *E.coli* network is summarized in Table 2. Topological analysis revealed that the global structures of the random networks still preserve a global bow-tie, but substantial difference exists between the *E.coli* metabolic network and randomized ones in term of the sizes of bow-tie parts and average clustering coefficient.

It can be seen that the sizes of GSC part and 2-core are much smaller than those of random networks respectively. More significantly, no 3-cores have been detected in random networks. The value of Z-score suggests that such difference is unlikely to arise at random. Table 2 also shows that the clustering coefficients of the random networks are almost equal to zero, in big contrast to that of the *E.coli* network. This clear difference implies an overall loose connection of randomized networks but a cliquish topology of the *E.coli* metabolic network. Such different topological patterns are observably presented in their 2-cores, as Figure 5 shows. The 2-core of *E.coli* network exhibits distinct cohesive areas being sparsely linked together, while the randomized one is linked in such an approximately equal density that almost no clear-cut "cliques" appear within it, especially, the random network doesn't have 3-cores. These comparative results indicate that the cliquish bow-tie topology is an intrinsic and significant feature of metabolic networks, rather than a random phenomenon. Such topology pattern could be the result of local interactions within metabolic pathways. It also agrees with earlier observations concerning the modular organization of metabolic networks [34, 36-38].

Table 2  Comparison of the *E.coli* metabolic network with sixty randomized networks

|  |  | GSC | S | P | IS | 2-core | 3-core | C |
|---|---|---|---|---|---|---|---|---|
| Sixty randomized networks | Mean | 287 | 90 | 126 | 71 | 205 | 0 | 0.0027 |
|  | Standard deviation | 15.86 | 10.23 | 14.37 | 13.72 | 12.43 | 0 | 0.0019 |
| *E.coli* network | | 234 | 85 | 177 | 79 | 163 | 28 | 0.0646 |
| Z-score | | -3.40 | -0.52 | 3.53 | 0.61 | -3.37 | ∞ | 31.91 |

C: Average clustering coefficient of the network

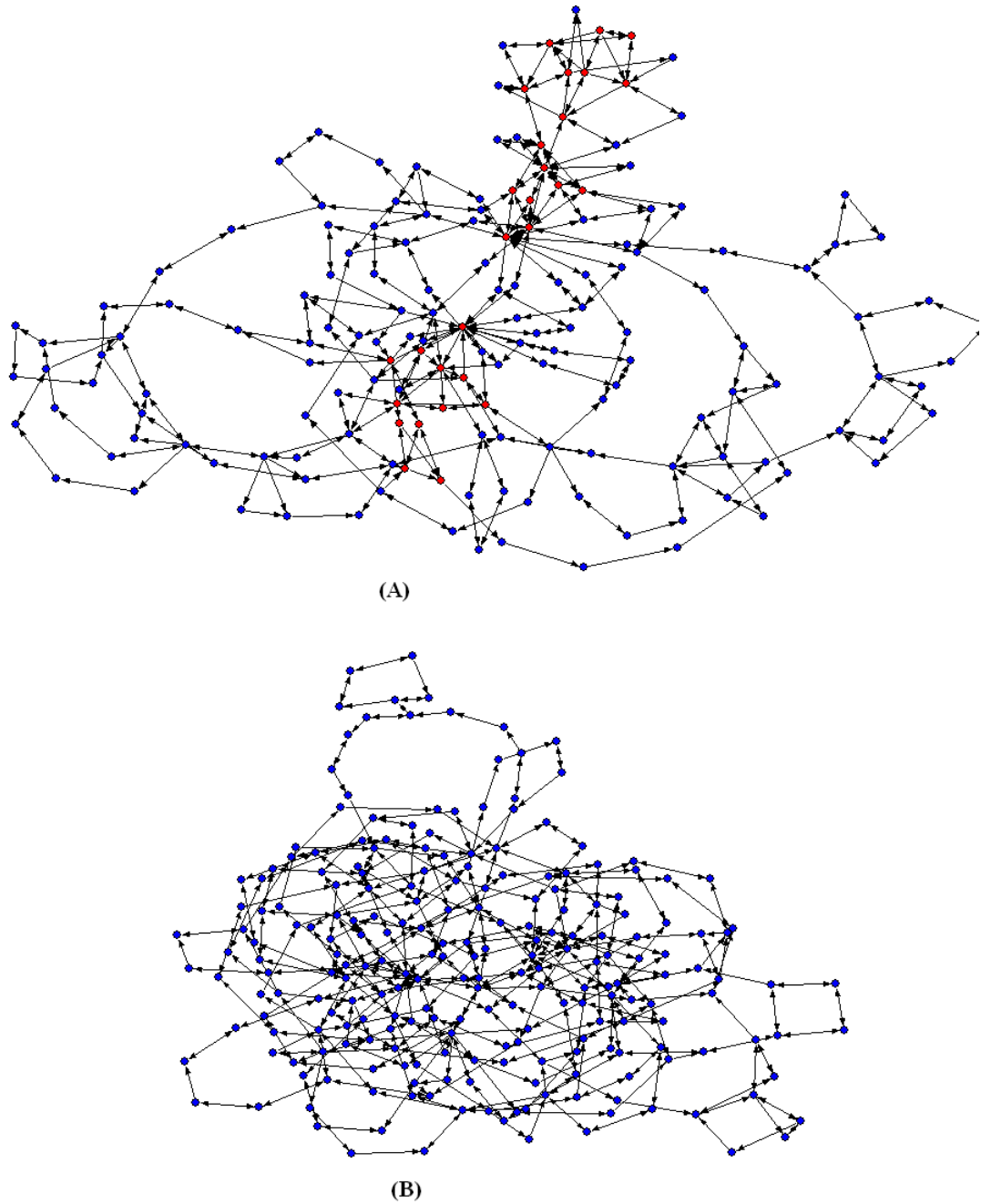

Fig. 5. Comparison of the 2-core of *E.coli* network with that of a randomized network.

(A) 2-core for *E.coli* network, 163 nodes included. The red nodes constitute 3-core.

(2) 2-core for a randomized network. The GSC, S, P and IS part include 302, 84, 134, 55 nodes respectively. The 2-core includes 227 nodes. It does not have a 3-core.

*2.5 Significance of bow-tie topology to metabolism*

    The topology of metabolic networks could be a reflection of the dynamics of their formation and evolution. Selection of bow-tie architecture seems to be a concise and smart option for constructing metabolic networks. From the standard biochemical point of view, the metabolic system is organized as a bow-tie whose knot is made up of a small handful of activated carriers and 12 precursors, with a large "fan-in" of nutrients, and a large "fan-out" of products in

biosynthetic pathways[20, 23]. Such organization pattern has been reported to be present in various biological systems, such as in signal transduction systems, transcription and translation processes, and immune systems [20, 23, 26, 39-41]. The bow-tie model here could give alternative view of the biological metabolite flow from the topological aspects, where the knot is much thicker than that of above. We will refer to thin bowtie and thick bowtie to distinguish these two models. These two bow-tie models are similar in that they both specially identify and isolate the carriers. It is noted that, besides the carriers, the thin bow-tie model includes only the 12 precursors as its knot, whereas the thick bow-tie model here also contains these 12 precursors, but together with the three essential pathways – TCA (tricarboxylic acid) cycle, pentose phosphate pathways and glycolysis pathways, which generate the precursors, as well as much more metabolites and reactions. Although different bow-tie models in details, the similar organization pattern can both facilitate the kind of extreme heterogeneity that allows for robust regulation, manageable genome sizes and biochemically plausible enzymes [23].

The knot in our model denotes the most tightly connected part of the network and is comprised of concentrated intermediated metabolites. This thicker knot would possibly allow the network to manipulate flexible controls through the knot and provide more interfaces with inputs and outputs to meet an emergency or process new metabolites. On the other hand, the thicker knot may reveal the flexibility that the organism has in interchanging nutrients and products. *E. coli* in particular heavily uses products of other organism metabolism as nutrients, as do most organisms, but can also live on fairly minimal media as well. The thick knot may reflect this flexibility, but further research will be needed to full explain these connections.

Another contribution would be network robustness. It is argued that the GSC part in the bow-tie of the metabolic network is robust against mutations because there are multiple routes between any pair of nodes within the GSC [15, 26]. More significantly, the fact that all the 12 precursors locate in the GSC, while half of them in the 3-core of *E.coli* metabolic network, the most densely connected part of the network, indicates a protective mechanism of metabolic networks. On the contrary, the elements outside the GSC and with a low degree of connectivity could probably be the most crucial ones for the entire system functioning. This argument also agrees with the results of Palumbo *et al* [10] and Samal *et al* [11] concerning the distribution of essential enzymes in the metabolic networks of *S. cerevisiae* and *E.coli,* respectively. Since the spread bow-tie model of this work also visualizes vulnerable connections in the network, i.e, elimination of a thin arc usually causes the splitting of the network, it may also be useful for investigating the vulnerability of metabolic networks or identifying drug targets.

## 3. Conclusion

This paper presents a spread bow-tie model based on rational reduction of complex metabolic network as continuous effort to explore the structure and functionality of large-scale metabolic networks. The spread bow-tie model provides better clarity of the overall structure and the biological information flow in the entangled metabolic pathways. It could be applied to study the vulnerability of metabolic networks. Further investigation to the reciprocal links and main cores in the GSC part suggests that the bow-tie of metabolic networks enriches bi-directed links in the GSC part and exhibits a cliquish character, whereas such features are not present in random graphs with comparable statistical weight. The results of this study afford a deeper comprehension of the design principles of metabolic networks.

**Acknowledgements** We thank H.W. Ma and A.P. Zeng for providing us with their metabolic networks' database. We also thank Dr. Lei Liu for critical reading of this paper. This work was supported in part by grants from Ministry of Science and Technology China (2003CB715900, 04BA711A21, 2004CB720103), the National Natural Science Foundation of China (Grant NO.30500107), and Science and Technology Commission of Shanghai Municipality（04DZ19850, 04DZ14005）.

579:5461-5465.